# A Differentiable Ray–Wave Framework for Hybrid Refractive–Diffractive System Modeling and Optimization

Jiazhou Cheng, Margaret Gao, Yixuan Shao, Chenkai Mao, Tom D. Milster, and Jonathan A. Fan*

**ABSTRACT:** Hybrid optical systems combining refractive and diffractive optical responses have the potential to support new types of optical behavior, but they are difficult to model and optimize due to the disparate spatial scales and physics exhibited by ray and wave phenomena. In this work, we present a differentiable ray–wave framework for modeling hybrid refractive–diffractive optical systems that operates as a plug-and-play module within standard ray tracing pipelines. Our model uniquely applies to both planar and curvilinear diffractive surfaces and accommodates arbitrary scalar holographic profiles with high spatial frequency responses, with each simulation evaluated at a single wavelength. We analyze ray–wave modeling regimes that optimally account for the spatial frequency properties and spatial curvature of the diffractive surfaces, and we demonstrate the gradient-based end-to-end optimization of hybrid refractive–diffractive systems featuring planar and conformal diffractive surfaces. We anticipate that these modeling capabilities will enable new classes of hybrid optical systems relevant to computational imaging and display applications.

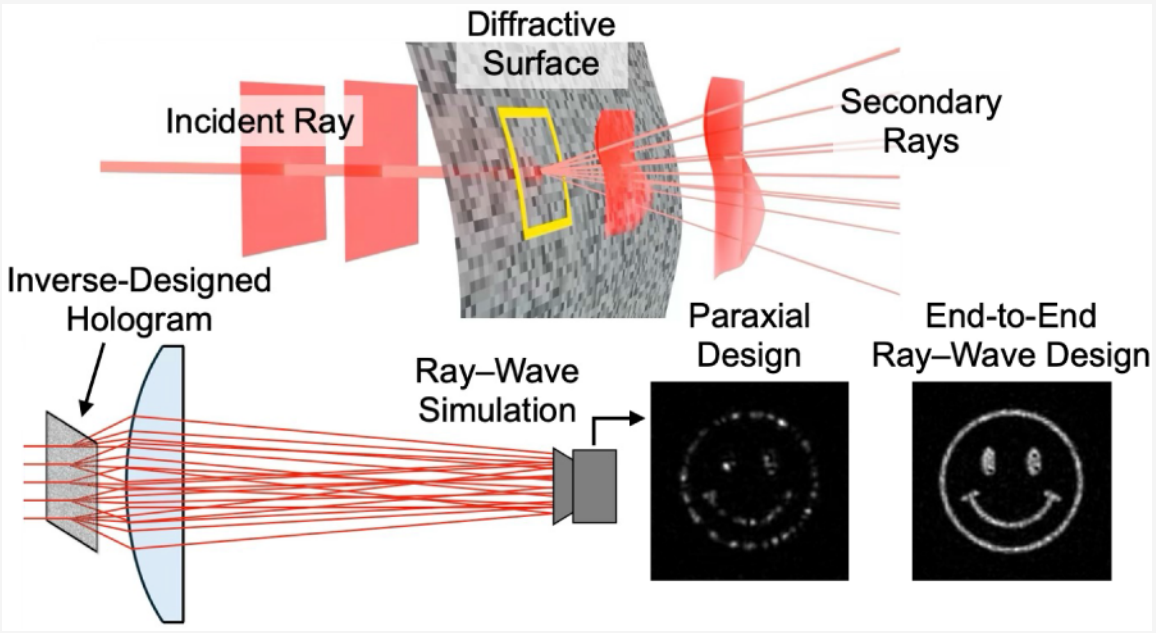

## INTRODUCTION

Hybrid optical systems that combine refractive surfaces with diffractive optical elements (DOEs) have the potential to enable optical systems with new form factors, including compact and conformal configurations, as well as new functionalities. They have emerged as a powerful platform for imaging, yielding systems featuring an ultrawide field of view (FOV)[1,2] and superachromatic responses.[3,4] They have also enabled new classes of nonimaging optics, including curvilinear free-space[5−7] and waveguide-based displays[8−10] for virtual and augmented reality systems, as well as freeform metasurface cloaks.[11−13] Such distinctive capabilities presented by hybrid optical systems are made possible by their ability to utilize the complementary physics of ray and wave optics. On the one hand, refractive ray-based optics support efficient, broadband, and field-dependent responses[14] with capabilities that have been enhanced by the development and implementation of freeform surfaces.[15] On the other hand, DOEs, including metasurfaces,[16] can support engineered wavefront responses[17,18] that exhibit polarization control,[19,20] tailored chromatic dispersion,[21] and angular and spectral selectivity.[22,23] Modern advancements in computational design and optimization[24−26] have pushed the capabilities of freeform metasurface DOEs to extreme physical limits.

To design hybrid refractive–diffractive optical systems, modeling tools capable of handling ray and wave optics are required (Figure 1a). These tools are challenging to develop because they must synergistically accommodate the distinctive physics of ray-based interactions, which are typically modeled using geometric ray tracing,[27] and wave-based physics, which is generally computed using Fourier optics (i.e., the angular spectrum method (ASM))[28] and numerical full-wave simulations.[29] To date, several modeling and design platforms for hybrid optical systems have been proposed and implemented, each applicable to specific classes of systems, but each with limitations that restrict their generalizability. Most hybrid simulators model DOEs based on the generalized law of refraction,[30] where an incoming ray produces a single outgoing ray with a direction determined by the local phase gradient. These concepts require the DOE phase profile to be locally smooth and do not apply to DOEs featuring complex-amplitude modulation. Nondifferentiable[31] and differentiable frameworks[32−34] for this form of ray–wave simulation have been implemented, with the latter made possible by specifying the DOE phase profile as an analytic and differentiable function. Other methods utilize ASM to model the wave response of DOE surfaces,[35,36] but they restrict the DOE to planar form factors and they require the DOE to have locally smooth phase gradients or to be the last element in the optical system. Hybrid simulators that can model arbitrary apertures or pinholes (i.e., amplitude DOEs), which are based on the Huygens−Fresnel principle or Rayleigh−Sommerfeld diffraction integral, have also been proposed[37,38] but are non-differentiable. A summary of the features of representative published hybrid simulators is in Supporting Information S1 and Table S1.

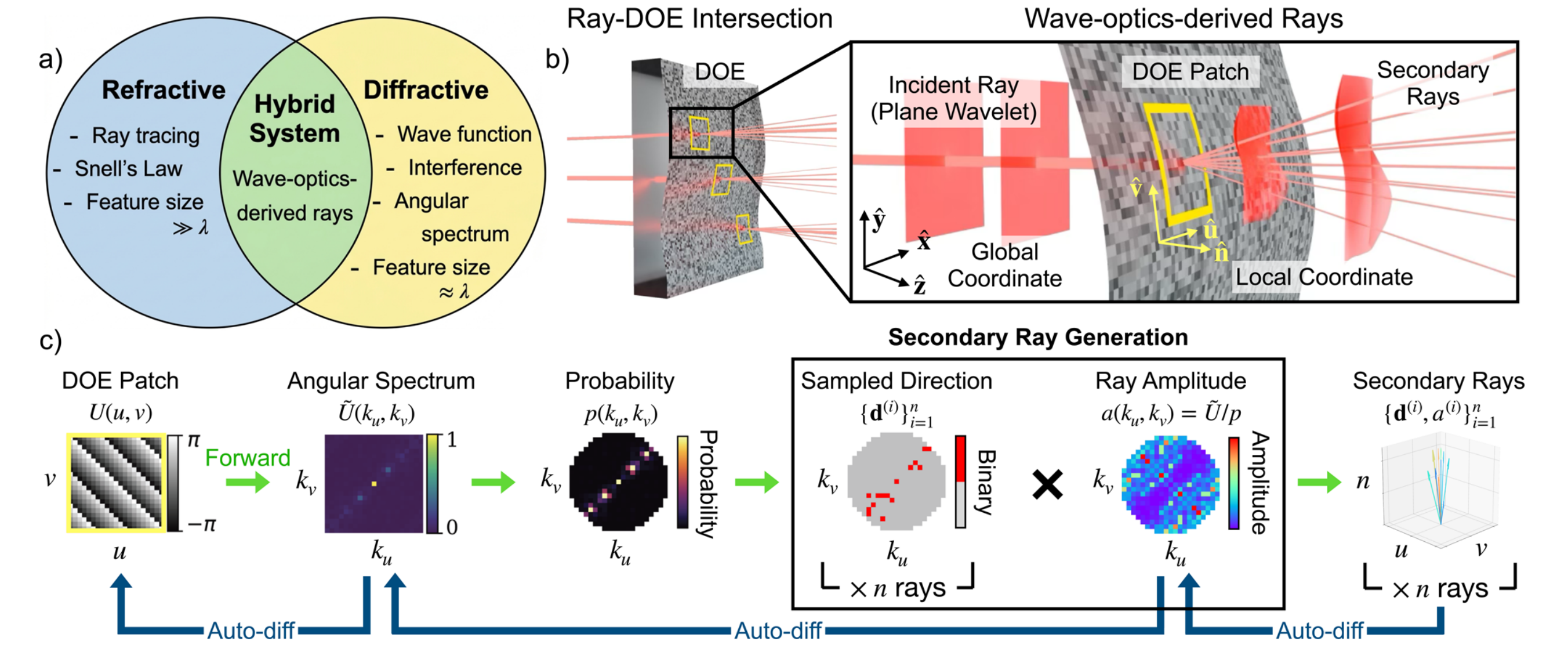

**Figure 1.** Differentiable ray−wave framework for hybrid refractive−diffractive optics. (a) Conceptual comparison between refractive ray physics, diffractive wave physics, and the hybrid regime addressed in this work. (b) Schematic of rays interacting with a diffractive optical element (DOE). At each ray−DOE intersection, a local DOE patch is extracted, and its far-field response is calculated using the angular spectrum method and sampled into secondary rays. (c) Sampling pipeline used for differentiable optimization. A DOE patch is Fourier transformed into a continuous angular spectrum $\tilde{U}(k_u, k_v)$, from which the continuous probability distribution $p(k_u, k_v)$ used to sample secondary rays is derived. In the forward pass (green arrow), $N$ discrete wavevectors are stochastically sampled as propagation directions of the secondary rays, and each secondary ray is assigned a complex-amplitude weight $a^{(i)} = \tilde{U}(k_u^{(i)}, k_v^{(i)})/p(k_u^{(i)}, k_v^{(i)})$. During backpropagation (blue arrow), gradients propagate through $\tilde{U}(k_u^{(i)}, k_v^{(i)})$ within $a^{(i)}$ to update the DOE profile, bypassing the nondifferentiable secondary ray sampling step.

In this work, we present a differentiable ray−wave hybrid modeling framework that supports the simulation and gradient-based inverse design of a wide range of hybrid refractive−diffractive systems. Our method is based on explicitly computing local DOE scattering responses using ASM, which are sampled back into rays using Monte Carlo estimation[38−40] coupled with a gradient estimator that maintains the differentiability of the wave-to-ray transformation. It enables coherent wave modeling of systems featuring arbitrary holographic DOE profiles, including those supporting complex-amplitude modulation and large wave-vector responses beyond the paraxial regime. It also accommodates the placement of DOEs anywhere within multielement optical systems and applies to systems featuring spatially curvilinear DOEs. Importantly, our platform operates as a plug-and-play module for ray−wave−ray DOE interactions that can readily integrate with existing ray tracer codes.

## ■ RESULTS AND DISCUSSION

### Ray−Wave Simulator Formalism

Broadly, our framework models hybrid optical systems by specifying geometric rays as plane-wave wavelets with defined amplitudes and phases. When an individual ray intersects a DOE surface, it interacts with a finite-sized DOE patch and produces a local diffractive response comprising an angular distribution of scattered wave vectors (Figure 1b). This distribution is sampled to yield an ensemble of secondary rays, each representing a plane-wave component of the local angular spectrum, and these secondary rays are then propagated using conventional ray-tracing principles. A sufficient number of incident rays are specified such that the ensemble of rays provides a uniform coverage of the DOE surface with patches. Our concept can be generalized to systems containing multiple DOEs (Supporting Information S4), but for our analysis here, we limit our discussion to systems containing a single DOE.

A more detailed treatment of the local diffractive response of an individual ray interacting with a DOE is shown in Figure 1c. We first construct a tangent coordinate frame $(\hat{\mathbf{u}}, \hat{\mathbf{v}}, \hat{\mathbf{n}})$ at the point of ray−DOE intersection. This intersection point specifies the center of the DOE patch, which is defined to have a square geometry with dimensions that are a design choice. The DOE patch can have a phase, amplitude, or complex-amplitude response, with the general form $U(u, v) = A(u, v)\exp(i\phi(u, v))$. The Fourier transform of this patch response yields its local scattering response $\tilde{U}(k_u, k_v)$ as obtained using the angular spectrum method, where $(k_u, k_v)$ denote transverse wave vectors in the tangent plane. Only propagating modes are retained, and evanescent components satisfying $k_u^2 + k_v^2 > k^2$ are discarded. When the fast Fourier transform is used for ASM calculation, the scattering response is discretized in wave vector space due to the discrete nature of the transform.

The scattering profile is subsequently converted into a set of secondary rays by using Monte Carlo sampling. The direction of each secondary ray, $(k_u^{(i)}, k_v^{(i)})$, is sampled from a density function $p(k_u, k_v)$, which relates to $\tilde{U}(k_u, k_v)$ in a manner that will be analyzed later. The complex amplitude specified for the $i$-th sampled ray, up to a global normalization factor, is given by

$$a^{(i)} = \frac{\tilde{U}(k_u^{(i)}, k_v^{(i)})}{p(k_u^{(i)}, k_v^{(i)})} \tag{1}$$

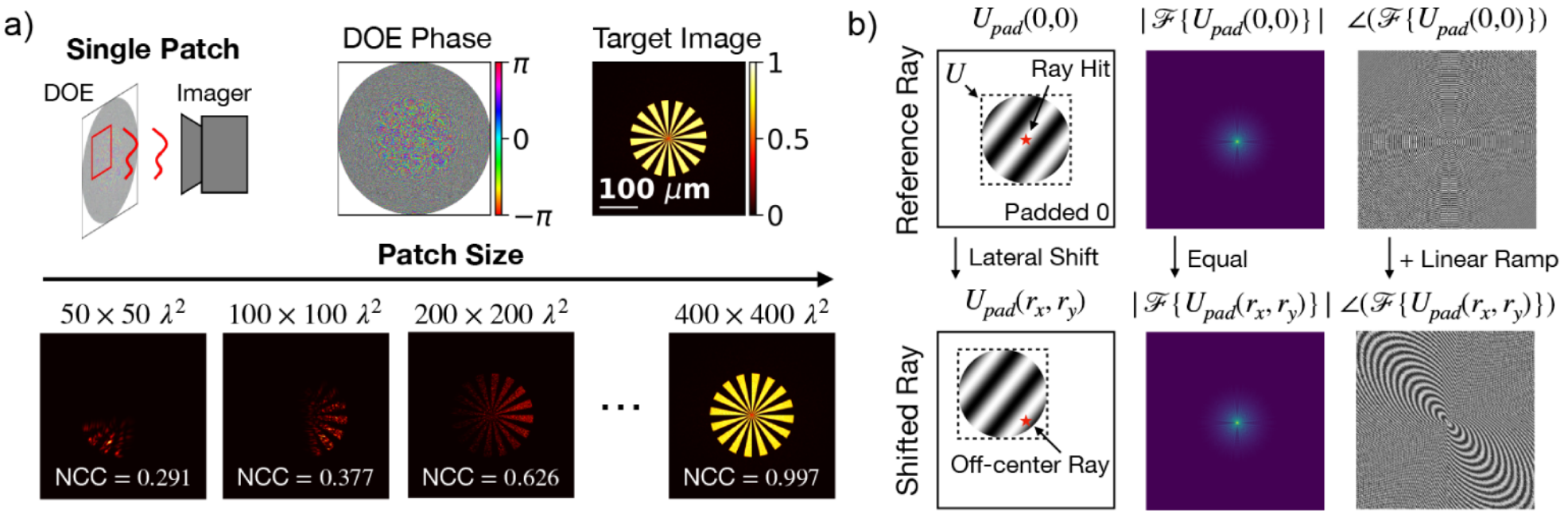

**Figure 2.** Ray–wave modeling in the single DOE patch limit. (a) Ray–wave modeling of a holographic target with a single DOE patch (size 400 × 400$\lambda^2$), which captures a limited local response of the diffraction element. As the patch size increases from 50 × 50$\lambda^2$ to 400 × 400$\lambda^2$, the reconstructed field from the patch progressively approaches the full-field result (bottom right). (b) Rule-of-thumb ray–wave modeling approach for planar DOEs. The full DOE field should be zero-padded to twice its original size, and its angular spectrum response for a centered incident ray can be obtained using a single FFT calculation. The response of corresponding off-center incident rays is readily obtained by applying a linear phase ramp relative to the on-center FFT response.

This secondary-ray complex amplitude is inversely proportional to $p(k_u, k_v)$, which accounts for the fact that rays with larger probability are sampled more frequently.[41] In the limit where an infinite number of sampled rays are considered, the secondary ray distribution converges to $\tilde{U}(k_u, k_v)$. As this secondary ray representation incorporates both local angular spectrum and sampling density information, it faithfully represents the local DOE response, and a proof of this equivalence is provided in Supporting Information S2.

After forward propagation through the optical system, each secondary ray is characterized by its spatial position on the sensor plane, $\mathbf{r}^{(i)}$, direction $\mathbf{d}^{(i)}$, complex amplitude $a^{(i)}$, and accumulated optical path length $\mathrm{OPL}^{(i)}$. The complex field is reconstructed coherently at the sensor plane by summing the rays as plane wave wavelets, which is consistent with protocols from other ray tracers, including ZEMAX:

$$U_{\mathrm{sensor}}(x, y) = \sum_i a^{(i)} \exp[ik(\mathrm{OPL}^{(i)} + \Delta r^{(i)}(x, y))]\langle \mathbf{n}, \mathbf{d}^{(i)} \rangle \tag{2}$$

In this expression, $k = 2\pi/\lambda$ is the free-space wavenumber, $\langle \mathbf{n}, \mathbf{d}^{(i)} \rangle$ represents the projection factor between the wavelet direction and the sensor-plane normal, and $\Delta r^{(i)}(x, y)$ accounts for the additional optical path differences between the ray's intersection point on the sensor plane and the field-evaluation point $U_{\mathrm{sensor}}(x, y)$ due to the oblique incidence of the wavelet across the sensor plane.

To perform backpropagation through our ray–wave model in a manner where the DOE profile is updated to improve the field response at the sensor, we use a gradient estimator technique (Figure 1c). In particular, to sidestep the fact that secondary ray sampling is inherently nondifferentiable, backpropagated gradients from the secondary rays flow through $\tilde{U}(k_u, k_v)$ in $a^{(i)}$ back to $U(u, v)$ for all secondary rays, while $p(k_u^{(i)}, k_v^{(i)})$ in $a^{(i)}$ is treated as constant in backpropagation. This estimator neglects gradients associated with changes in the sampled secondary-ray directions, but it nonetheless provides low-variance gradients that are sufficient for stable DOE optimization,[42,43] as demonstrated by the inverse design results in later sections. Detailed implementation and alternative differentiable sampling strategies are described in Supporting Information S7.

Accurate implementation of the ray–wave framework requires careful selection of the simulation parameters, including the sampling density of incident and secondary rays and the DOE patch size. We subsequently perform a systematic analysis of these hyperparameters, starting with an examination of patch size. The optimal choice of patch size depends strongly on whether the DOE is planar or curvilinear, and on whether the full DOE field can be represented within GPU memory. In Figure 2a, we simulate a 400$\lambda$ × 400$\lambda$ planar holographic system with a single incident ray. We initially consider $D \times D$ patch areas for each ray with $D$ = 50, 100, 200, and 400$\lambda$. The hologram has an NA = 0.7 and is designed for an operating wavelength of $\lambda$ = 1 $\mu$m. Pattern formation at the imager focal plane is computed by propagating each DOE patch via a sufficiently large number of sampled secondary rays (SSRs) to ensure convergence of each patch and coherently summing all the ray contributions at the imaging plane. To calculate the angular spectrum, all patches are zero-padded to a total patch area of 400 × 400$\lambda^2$ to ensure the angular spectrum resolution is high for all patches and to avoid the presence of spurious periodic copies of the scattered field in the reconstructed sensor field. To quantify the fidelity of image reconstruction at the imager focal plane, we calculate the normalized cross-correlation (NCC) with respect to the ASM ground-truth reference $I_{\mathrm{GT}}$:

$$\mathrm{NCC}(I) = \frac{\sum_{x,y} I_{\mathrm{GT}}(x, y) I(x, y)}{\sqrt{\sum_{x,y} I_{\mathrm{GT}}(x, y)^2 \sum_{x,y} I(x, y)^2}} \tag{3}$$

We observe that when a single ray is incident on the hologram, the use of larger DOE patch sizes yields higher NCC values at the sensor plane, as larger patches better sample the total DOE profile (Figure 2a, bottom).

On the basis of these observations, we propose the following rule-of-thumb modeling method for planar DOEs. We first zero-padded the DOE field to twice the lateral extent of the original phase profile, such that the full DOE response is captured within this field for all incident rays. We then computed the FFT of the global zero-padded DOE field, which corresponds to the DOE response of a normal, centered ray. The FFT responses of nonnormal, off-center incident rays are then readily calculated analytically as the addition of linear phase ramps to the original angular spectrum response, as specified by the Fourier shift theorem. In this manner, only a single FFT calculation is performed, independent of the number of incident rays. The incident ray count is determined by sampling requirements in the downstream optical system, including pupil coverage, field-dependent aberrations, vignetting, and subsequent optical surfaces.

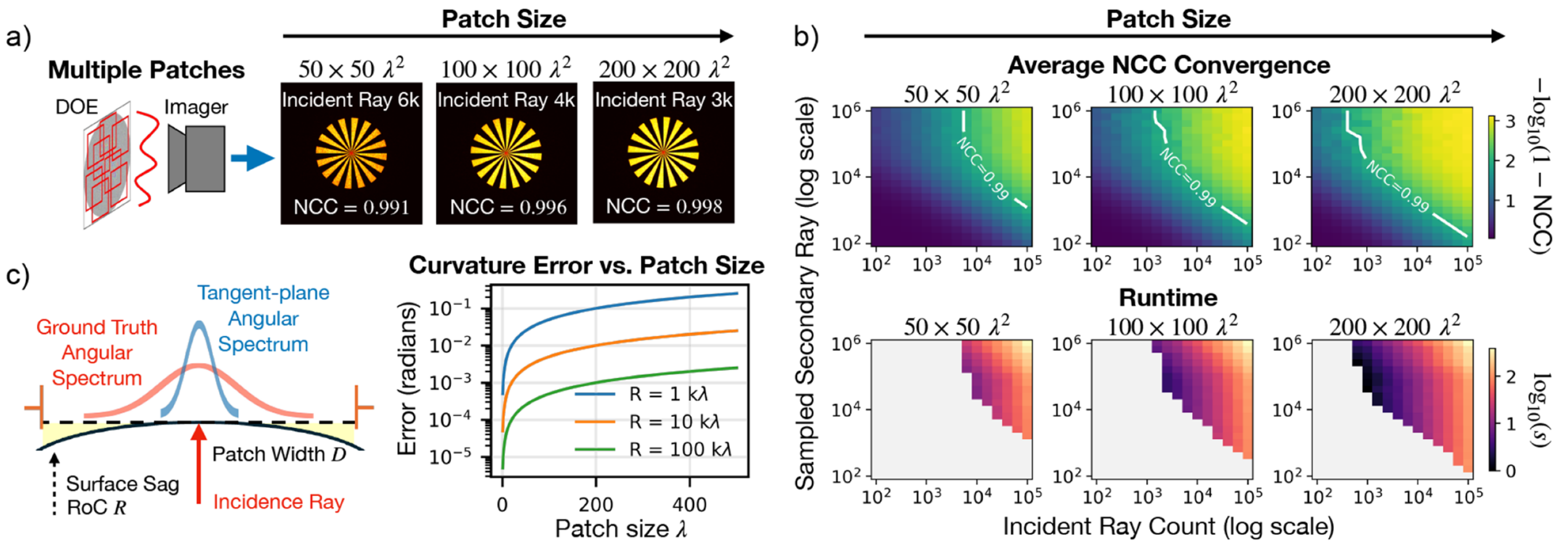

**Figure 3.** Field convergence of multiple overlapping patches and trade-offs in ray–wave modeling. (a) Ray–wave modeling of a holographic target with multiple patches. The DOE is sampled with different numbers of incident rays and the same sufficiently large number of SSRs for different patch sizes. The total DOE optical response is calculated by summing the diffractive response of all rays. All patch sizes achieve high-fidelity reconstructions with NCC >0.99, and larger patches require fewer incident rays. (b) Reconstruction accuracy and runtime with respect to incident ray density and SSR density for different patch sizes. Each condition is averaged over 4 trials. The top row shows $-\log_{10}(1 - \text{NCC})$, with white contours indicating NCC = 0.99. Larger patch sizes shift this contour toward lower total ray counts. The bottom row shows the runtime for simulations achieving NCC ≥0.99. The corresponding peak GPU memory consumption for each condition is provided in Supporting Information S5 and Figure S2. (c) Curvature-induced error for conformal DOEs, which arises when flat DOE patches are used to approximate curvilinear DOE patches. Schematic (left) of the tangent-plane approximation on a curved surface and corresponding angular spectrum error (right) as a function of DOE patch size for different curvilinear DOE radii of curvature. This analysis applies to all phase profiles.

There are practical cases where the full-field planar-patch approach is not feasible. For large planar DOEs where the processing of the full-field planar patch exceeds available GPU memory, we can instead decompose the full field into multiple smaller overlapping patches, each associated with an incident ray. Using the same free-space hologram setting in Figure 2a and three patch sizes $D$ = 50, 100, 200$\lambda$, we consider an ensemble of incident rays with collectively overlapping DOE patches to model the total DOE response. When a sufficient number of incident rays are used, the coherent superposition of the transmitted fields at the sensor plane can accurately reconstruct the full field for all $D$ (Figure 3a). To further quantify the relationship between DOE patch size, incident ray density, SSR density, and reconstruction accuracy at the sensor plane, we sweep the incident-ray count from $10^2$ to $10^5$ and the SSR count per patch from $10^2$ to $10^6$ for the representative patch sizes. In Figure 3b, the top row shows reconstruction fidelity as $-\log_{10}(1 - \text{NCC})$, with the white contour marking NCC = 0.99, which specifies the spatial and angular sampling combinations sufficient for accurate reconstruction. NCC increases with both the incident-ray count and the SSR count, which are related to DOE spatial coverage and angular-spectrum sampling convergence, respectively. Larger patches shift the contour to lower total ray counts, thereby reducing the number of overlapping patches required and the corresponding runtime (Figure 3b, bottom row). Based on these observations, we find that the most computationally efficient modeling method involves specifying the largest patch size possible that fits within the available GPU memory.

For conformal, curvilinear DOEs, proper selection of DOE patch size strongly depends on the surface curvature of the DOE element. In particular, our method computes the local field response from a DOE patch using a tangent coordinate frame that approximates the surface as locally planar, and this approximation can lead to distortions in the local angular spectrum response (Figure 3c, left). These distortions are amplified when the patch size becomes larger or the DOE element curvature gets smaller: in both cases, phase error increases due to enhanced sag. We derive an upper bound on this curvature-induced error in the angular spectrum for a DOE element with radius of curvature ($R$):

$$\varepsilon_{\text{curv}} \leq \arcsin\left(\frac{D}{2R}\right) \tag{4}$$

This bound is independent of the detailed DOE phase profile. Details of this derivation can be found in Supporting Information S3. A plot of angular spectrum error as a function of DOE patch size for different $R$ (Figure 3c, right) confirms that $\varepsilon_{\text{curv}}$ increases as $D$ increases and $R$ decreases. The optimal choice of $D$ for a given curvilinear DOE element is the largest possible value given a user's threshold for angular spectrum error, which balances computational memory with curvature-induced errors.

We next identify a protocol for specifying the sampling distribution and number of SSRs for a given ASM response from a DOE patch in a manner that yields accurate sensor-plane reconstruction. We consider the diffractive response from two DOE patch types, a 50 × 50$\lambda^2$ metalens phase profile with an NA = 0.1 and a 400 × 400$\lambda^2$ holographic Siemens star with an NA = 0.7. In both cases, a single input ray is launched at the center of the DOE patch, and secondary rays are sampled from a probability density function $p(k_u, k_v)$. For each patch, we consider a probability density function with uniform wavevector sampling, $p_{\mathrm{uni}}(k_u, k_v)$, and a probability density function with sampling that relates to the ASM scattering magnitude, $p_{\mathrm{mag}}(k_u, k_v) \propto |\tilde{U}(k_u, k_v)|$. We sweep the number of SSRs and evaluate the mean squared error ($\epsilon_{\mathrm{MSE}}$) relative to an ASM reference to compare convergence rates.

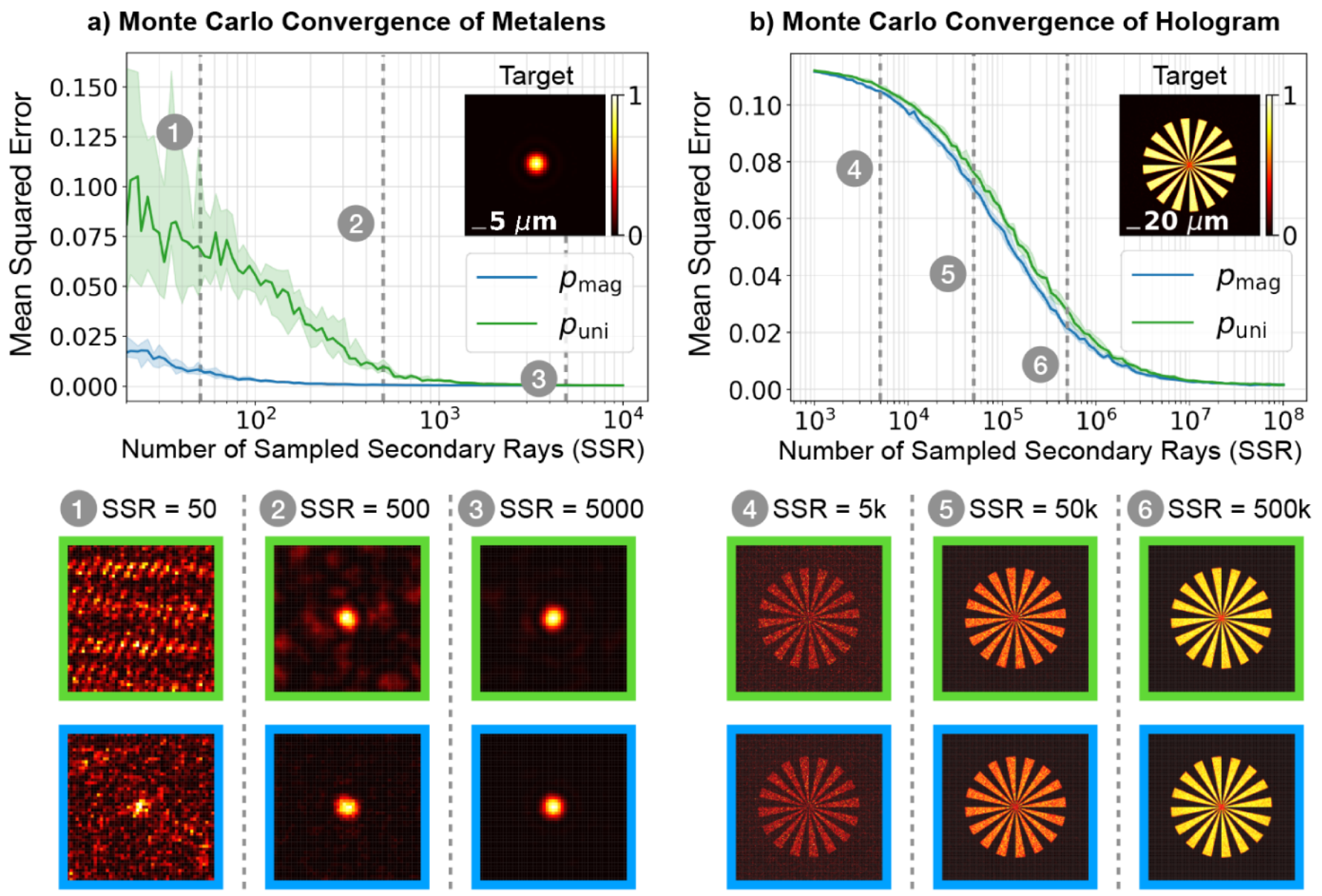

**Figure 4.** Monte Carlo convergence of sampled secondary rays (SSRs) per patch. (a) Mean squared error versus the number of SSR for a metalens DOE phase patch under uniform and spectral-magnitude-based sampling, with representative reconstructions at selected SSR values. (b) Corresponding convergence for a holographic DOE phase patch generating a Siemens-star pattern.

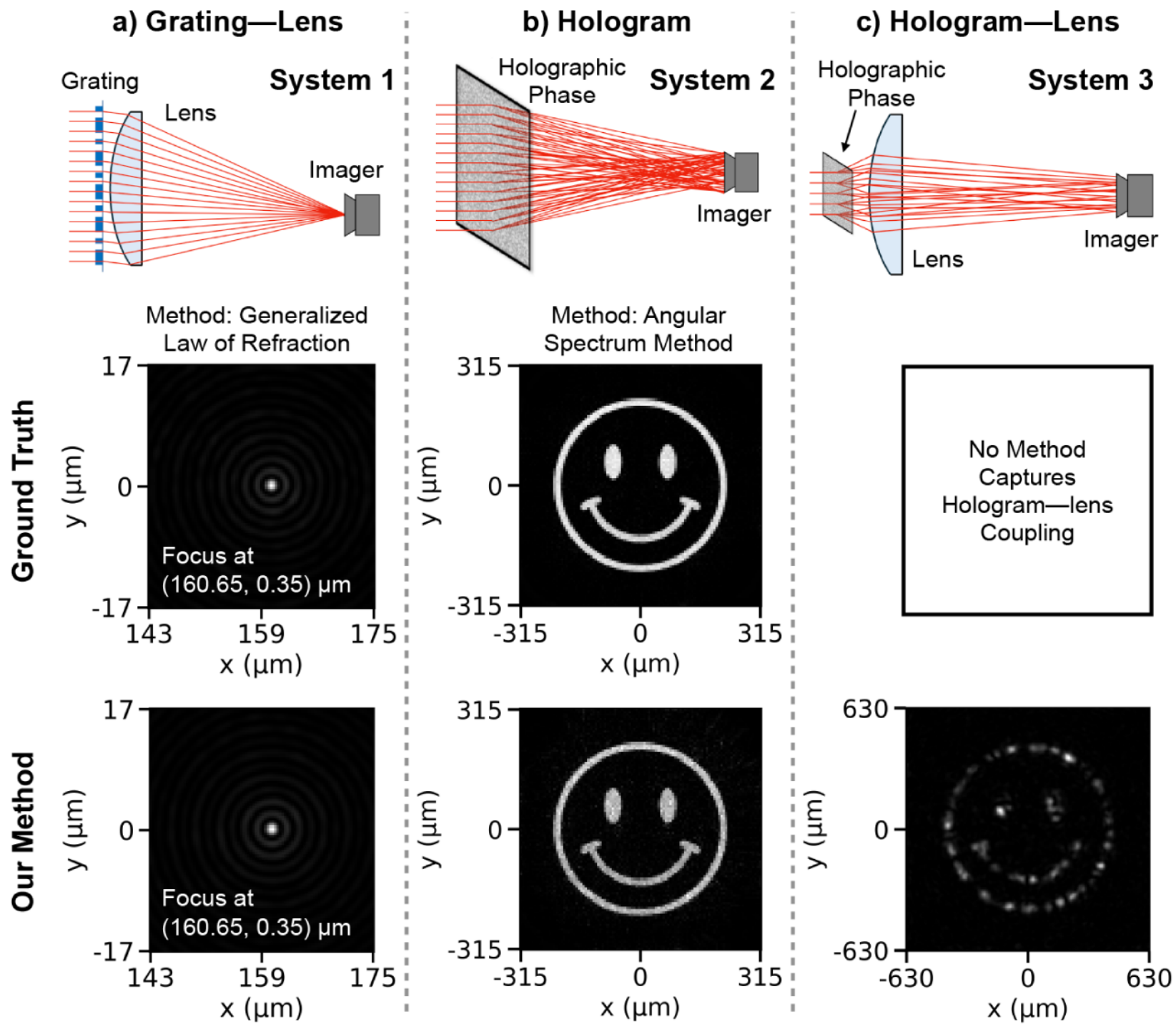

**Figure 5.** Benchmarking the simulation capabilities of our ray−wave tracer. We consider three model systems: (a) a grating−refractive lens pair, (b) a planar holographic surface, and (c) a planar holographic surface with a refractive lens. The top row shows schematics of the model systems. The middle row shows ground-truth images at the imaging plane calculated using the generalized law of refraction for (a) and the angular spectrum method for (b). No conventional simulation method exists for (c). The bottom row shows corresponding results from our ray−wave framework. The method reproduces the reference results in Setups 1 and 2 and predicts the nonparaxial response of Setup 3.

As shown in Figure 4, the reconstructed field converges to the ground truth solution as the number of SSRs increases for both phase profiles, indicating that high accuracy is always possible with sufficient SSR sampling. For the holographic Siemens star, which has a complex, multilobed diffractive response, the convergence rates for $p_{\mathrm{uni}}(k_u, k_v)$ and $p_{\mathrm{mag}}(k_u, k_v)$ are similar. For the metalens phase profile, which, in contrast, has an angular spectrum more concentrated in a single lobe, $p_{mag}(k_u, k_v)$-based sampling leads to accelerated convergence compared to $p_{uni}(k_u, k_v)$-based sampling. These demonstrations suggest that sampling proportional to the spectral magnitude provides a robust and generally effective strategy across different phase profiles.

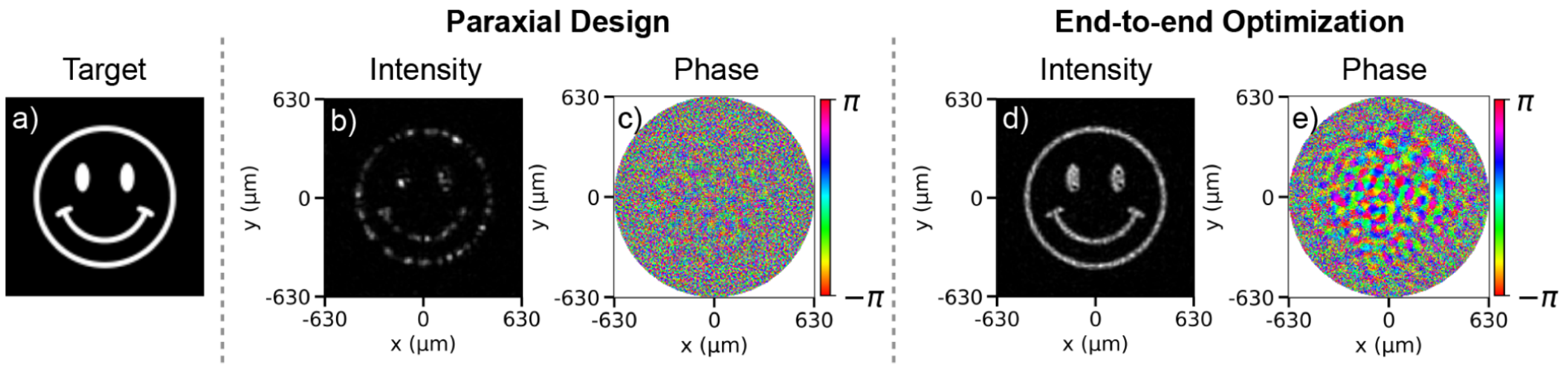

Figure 6. End-to-end optimization of a hybrid refractive−diffractive system. (a) Target intensity. (b,c) Sensor-plane intensity and DOE phase profile obtained from a paraxial design. (d,e) Sensor-plane intensity and DOE phase obtained from end-to-end optimization with our differentiable ray−wave model.

## Benchmark and Inverse Design Demonstrations

To benchmark the capability and accuracy of our ray−wave simulation platform, we implement it as a plug-and-play diffractive surface module within the open-source ray tracer DeepLens,[44] analyze three representative optical refractive−DOE systems, and compare our results with conventional methods. All simulations are performed with a wavelength of 0.7 $\mu$m. A summary of the computational settings, runtime, and GPU memory consumption for each simulation is provided in Supporting Information S8 and Table S2.

The first system (Figure 5a) is a planar diffractive grating placed 3 mm in front of a plano-convex lens (Thorlabs LA1131-A, diameter 25.4 mm, radius of curvature 25.8 mm, center thickness 5.3 mm, material N-BK7). The sensor plane is located 50 mm after the lens. The diffractive grating is assumed to possess an ideal linear phase profile, yielding a single diffraction order. Such a system can be readily modeled using the generalized law of refraction to describe the DOE, and analysis in this manner, which we consider to be the ground-truth reference, yields an aberrated focal spot at the sensor plane with a centroid position (160.65, 0.35) $\mu$m. With our method, we implement the linear phase profile on a discretized grid with a pixel pitch of 6.3 $\mu$m × 6.3 $\mu$m, which corresponds to the resolution of an off-the-shelf spatial light modulator (EXULUS-SE1, operating range 400 − 850 nm). A total of 6.5 × $10^4$ incident rays are sampled at the entrance pupil, and at each ray−DOE intersection, we adopt a patch-based ray−wave implementation (Figure 3) with a DOE patch size of 40 × 40 pixels (252 × 252 $\mu m^2$) to compute the angular spectrum, from which $10^3$ SSRs are generated and propagated to the sensor plane via conventional ray tracing. The ray sampling strategy and DOE patch size here and in the other demonstrations are determined based on the analysis in the previous section. The resulting intensity distribution of our method at the sensor plane closely matches the ground-truth focal spot. We obtain an MSE = 1.456 × $10^{-9}$ and an NCC = 0.985 with respect to the ground-truth intensity, indicating good agreement in both absolute intensity and structural similarity. The discrepancy between our simulated results and the ground truth arises in part from slight differences in the phase profiles used for the analysis (i.e., ideal linear versus discretized staircase profiles).

The second system (Figure 5b) consists of a planar holographic phase profile in free space that produces a smiley face image at the sensor plane. The phase profile is implemented using the same SLM configuration as in the first system (i.e., a pixel pitch of 6.3 $\mu$m × 6.3 $\mu$m and a total of 100 × 100 pixels). The sensor sampling pitch and total size are chosen to match those of the SLM, and the total system has an NA = 0.5. A ground-truth sensor-plane image is obtained using ASM. Using our framework, we use a single patch that spans the full field (Figure 2b), and one primary ray together with 1.1 × $10^6$ SSRs is sufficient to represent the full DOE response. We obtain an MSE = 4.414 × $10^{-10}$ and an NCC = 0.997, indicating good agreement between our method and the ground truth.

In the third system (Figure 5c), we construct a hybrid optical setup by cascading a planar hologram with the refractive lens (Thorlabs LA1131-A) to produce the smiley-face image at the sensor plane. Our specification of high-frequency wavefront modulation followed by conventional refractive propagation is typical in many computational imaging and display applications.[45] The holographic phase profile is defined on a 200 × 200 pixel grid with a sampling pitch of 6.3 $\mu$m, matching SLM specifications. There does not exist a conventional approach for precisely designing such a hybrid system, and we instead design the DOE in the paraxial pupil-phase approximation by replacing the refractive lens with an equivalent flat metalens. We apply our ray−wave simulation framework to the designed DOE surface together with the refractive lens, adopting the full-field patch rule (Figure 2b). Unlike the second system of a free-space hologram, an ensemble of incident rays is still required because the downstream refractive lens introduces field-dependent, non-paraxial aberrations. We therefore uniformly sample 3.3 × $10^4$ incident rays across the entrance pupil, and each patch is sampled into 1.6 × $10^5$ SSRs. We observe that the sensor intensity distribution preserves the global structure of the smiley face; however, the image exhibits pronounced speckles. As verified by the convergence analysis in Supporting Information S6 and Figure S3, the resolved interference structure is reproducible across independent Monte Carlo realizations, thereby confirming that the speckles are physical interference effects arising from ray−lens interactions in the nonparaxial limit rather than numerical artifacts. These speckles, visualized in Figure 5c, highlight a fundamental limitation in conventional design workflows: refractive optics are typically approximated to have a fixed paraxial phase, presenting an oversimplified optical propagation picture, particularly for coherent optical systems.

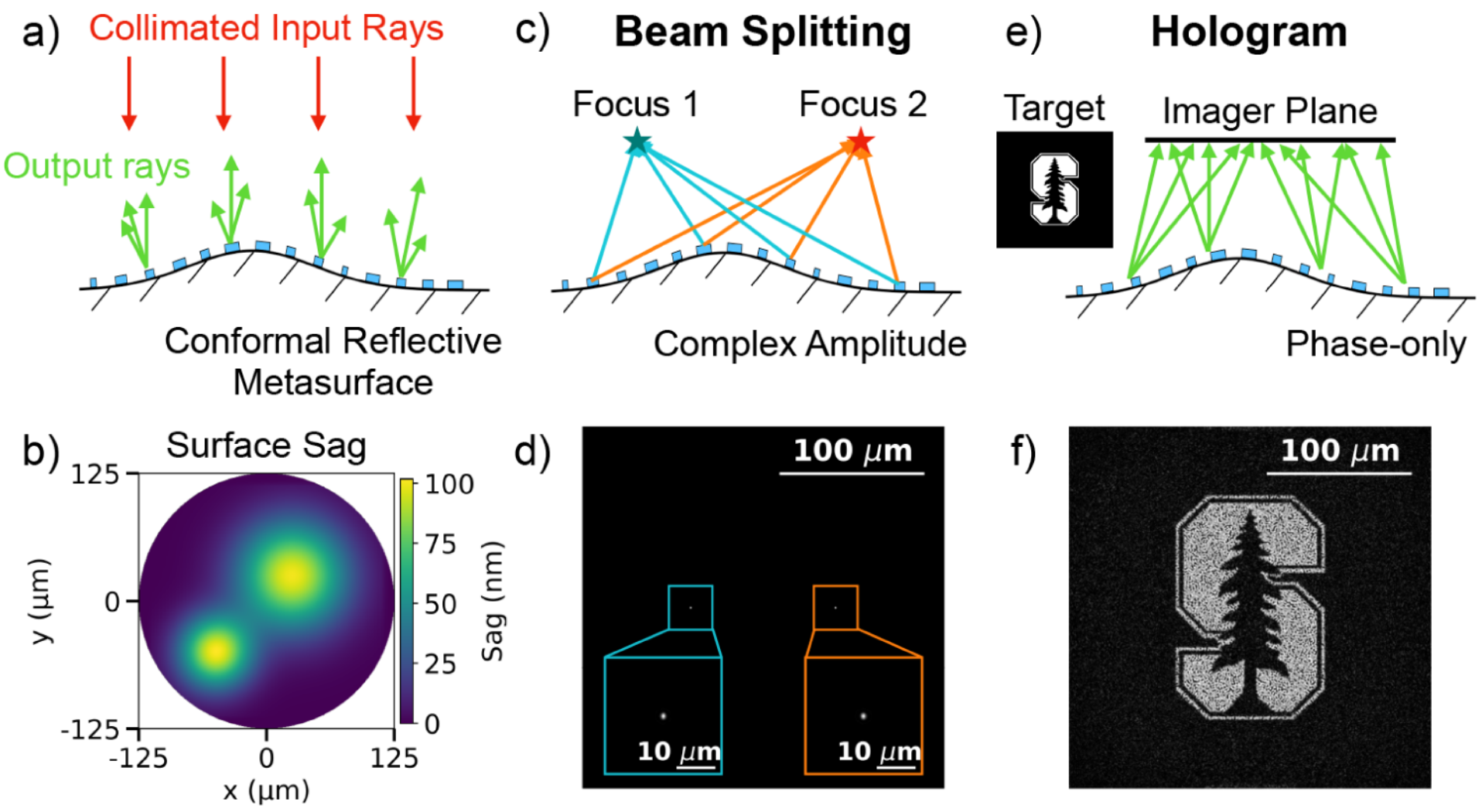

**Figure 7.** Optimization of a conformal reflective DOE on a curved substrate. (a) Schematic of the conformal reflective DOE under collimated illumination. (b) Surface sag of the curved substrate. (c,d) Beam-splitting task and optimized two-focus result. (e,f) Phase-only hologram task and reconstructed Stanford "S" image.

To address the design limitations mentioned above, we can utilize our differentiable ray–wave framework to directly perform end-to-end optimization of full DOE–refractive systems. A summary of computational settings, runtime, and GPU memory consumption for each design demonstration is provided in Supporting Information S8 and Table S3; additional details are provided in Supporting Information S9. Given a DOE phase profile $\phi(x, y)$, our method can readily simulate the coherent sensor field $S(\phi)$ and backpropagate gradients to update $\phi$. To minimize the discrepancy between the reconstructed intensity $|S(\phi)|^2$ and a target distribution $I_{tgt}$, we consider a normalized least-squares optimization objective:

$$\min_{\phi} \left\| \alpha |S(\phi)|^2 - I_{tgt} \right\|_2^2 \tag{5}$$

where $\alpha$ is a scalar normalization factor that compensates for global intensity scaling. The optimization results are shown in Figure 6. Compared with the paraxial design of System 3, our end-to-end DOE-refractive lens system fully accounts for aberrations and nonparaxial ray interactions introduced by the refractive lens, producing a qualitatively improved image at the sensor plane with high contrast and suppressed speckle artifacts. Quantitatively, the NCC improves from 0.462 with the approximate design method to 0.952 with the back-propagation-based design, indicating a substantial improvement in design fidelity.

Finally, we show that our ray–wave simulation platform can be used to inverse design the phase profiles of curvilinear DOEs. For these demonstrations, we consider a reflective metasurface DOE defined over a circular aperture of 501 × 501 pixels with a pixel pitch $\lambda/2$ and $\lambda = 1$ $\mu$m. The system has an NA = 0.5 and is illuminated by a collimated beam incident along the optical axis, which is normal to the global, flat $xy$-plane of the device (Figure 7a). The substrate possesses a physical Gaussian-like sag profile, as shown in Figure 7b. We use $3 \times 10^3$ incident rays that are uniformly distributed over the DOE and $4 \times 10^3$ SSRs per patch. A DOE patch size with $D = 100\ \lambda$ is selected based on our curvature error analysis (Figure 3c) and ensures that the error threshold $\varepsilon_{curv}$ is within 0.1 rad, given the minimum substrate radius of curvature of ≈ 506 $\lambda$. Inverse design is performed by backpropagation using eq 5.

We consider two design demonstrations. In the first demonstration, the conformal DOE is optimized to split the incident beam into two focused spots separated by 0.1 mm along the $x$-axis (Figure 7c). We specify the DOE to have a complex-amplitude field, as projected to a flat $xy$-plane, to be $U(x, y) = e^{i\phi_1(x,y)} + e^{i\phi_2(x,y)}$, and we optimize $\phi_1$, $\phi_2$. The optimized image at the sensor plane (Figure 7d) shows two sharp foci with a centroid separation of 99.99 $\mu$m and a mean full width at half-maximum (fwhm) of 1.2 $\mu$m (Supporting Information Figure S8), which is well within the 2.44 $\mu$m Airy disk diameter for this NA = 0.5 system. In the second demonstration, the DOE is defined as a phase profile $\phi(x, y)$ and tasked with generating a holographic Stanford "S" intensity distribution at the sensor plane (Figure 7e). Despite the phase perturbations introduced by the curved substrate, the optimization converges to a reconstructed image with an NCC = 0.909 (Figure 7f). In both cases, the optimized DOE profiles successfully encode the target diffraction patterns while simultaneously compensating for the curved substrate geometry.

## ■ CONCLUSION

We presented a differentiable ray–wave framework for the simulation and inverse design of hybrid optical systems containing refractive and diffractive elements. By representing ray–DOE interactions through local DOE patches, angular-spectrum-based secondary ray sampling, and coherent wavelet reconstruction at the sensor, the framework unifies local diffraction, global ray propagation, wave interference, and optical aberrations within a single differentiable simulator. We validated the method through convergence analysis, benchmark comparisons, and inverse design examples spanning planar DOEs with smooth and complex holographic profiles, hybrid DOE–refractive lens systems, and reflective conformal DOEs, showing that the framework can accurately model regimes that are difficult or impossible to capture with either conventional ray optics or purely wave-propagation methods alone.

Future work will focus on improving computational efficiency through adaptive patch sizing and adaptive ray sampling, enabling simulations that dynamically balance local diffraction accuracy and global ray propagation efficiency. We also envision efforts to (1) extend the framework to vectorial ray tracing, in which rays carry Jones vectors to model polarization-dependent propagation and diffractive interactions; and (2) couple the framework with full-wave solvers to rigorously model scattering beyond ideal complex-amplitude profiles, including wavelength-dependent resonances, dispersion, and anisotropic material responses. As the complexity of hybrid refractive−DOE systems increases with the advancement of fabrication technologies to produce freeform refractive surfaces, multilayer metasurfaces, and conformal metasurfaces, our proposed tools will become even more essential to enable accurate and robust system design. Along these efforts, we anticipate the need to further develop and utilize new full-wave simulation techniques, including those based on machine learning,[46−50] to manage the speed and memory requirements of large-scale full-wave calculations.

## ■ ASSOCIATED CONTENT

### Data Availability Statement

**Code Availability** The code for the ray−wave framework and simulation examples can be found at: https://github.com/jiazhou-cheng/raywave-tracing.

### Supporting Information

The Supporting Information is available free of charge at https://pubs.acs.org/doi/10.1021/acsphotonics.6c00818.

> Comparison of existing ray−wave frameworks; derivation of secondary-ray sampling and planar patch representation; curvature-induced error analysis; cascaded planar DOE propagation; Monte Carlo convergence analysis of the hologram−lens simulation; differentiable gradient estimation; computational cost and scalability; inverse-design details and additional results (PDF)

## ■ AUTHOR INFORMATION

### Corresponding Author

**Jonathan A. Fan** − *Department of Electrical Engineering, Stanford University, Stanford, California 94305, United States*; orcid.org/0000-0001-9816-9979; Email: jonfan@stanford.edu

### Authors

**Jiazhou Cheng** − *Department of Electrical Engineering, Stanford University, Stanford, California 94305, United States*; orcid.org/0009-0002-0094-9518

**Margaret Gao** − *Department of Electrical Engineering, Stanford University, Stanford, California 94305, United States*

**Yixuan Shao** − *Department of Electrical Engineering, Stanford University, Stanford, California 94305, United States*; orcid.org/0009-0009-6130-2831

**Chenkai Mao** − *Department of Electrical Engineering, Stanford University, Stanford, California 94305, United States*

**Tom D. Milster** − *James C. Wyant College of Optical Sciences, The University of Arizona, Tucson, Arizona 85721, United States*

Complete contact information is available at: https://pubs.acs.org/10.1021/acsphotonics.6c00818

### Funding

This work was supported by the National Science Foundation under Award No. 2103301 and the Air Force Office of Scientific Research under Award No. FA9550−25−1−0373. J. C. and M. G. acknowledge support from the Stanford Graduate Fellowship.

### Notes

**Associated Content:** Jiazhou Cheng; Margaret Gao; Yixuan Shao; Chenkai Mao; Tom D. Milster; Jonathan A. Fan. *A General Differentiable Ray−Wave Framework for Hybrid Refractive−Diffractive System Modeling and Optimization*. 2026, 2605.15418. arXiv. https://arxiv.org/abs/2605.15418 (accessed July 17, 2026).

The authors declare no competing financial interest.

## ■ REFERENCES

(1) Ma, X.; Wang, L.; Wang, X.; Song, W.; Sun, Q.; Zhu, L.; Huang, H. Learning Refractive-Diffractive Optics with Unidirectional Transformer for Large Field-of-View Imaging. *Int. J. Comput. Vis.* **2025**, *133*, 7570−7590.

(2) Park, Y.; Kim, Y.; Kim, C.; Lee, G.-Y.; Choi, H.; Choi, T.; Jeong, Y.; Lee, B. End-to-End Optimization of Metalens for Broadband and Wide-Angle Imaging. *Adv. Opt. Mater.* **2025**, *13*, 2402853.

(3) Stone, T.; George, N. Hybrid diffractive-refractive lenses and achromats. *Appl. Opt.* **1988**, *27*, 2960−2971.

(4) Flores, A.; Wang, M. R.; Yang, J. J. Achromatic hybrid refractive-diffractive lens with extended depth of focus. *Appl. Opt.* **2004**, *43*, 5618−5630.

(5) Nikolov, D. K.; Bauer, A.; Cheng, F.; Kato, H.; Vamivakas, A. N.; Rolland, J. P. Metaform optics: Bridging nanophotonics and freeform optics. *Sci. Adv.* **2021**, *7* (18), No. eabe5112.

(6) Wang, Y.; Zhang, C.; Meng, J. *Curved CGH design techniques for aligning reflective optical system*; SPIE, 2024.

(7) de Angelis, V.; Dorrah, A.; Ambrosio, L.; Capasso, F. Conformal holography with curved light sheets. *Opt. Express* **2025**, *33*, 6567−6580.

(8) Gopakumar, M.; Lee, G.-Y.; Choi, S.; Chao, B.; Peng, Y.; Kim, J.; Wetzstein, G. Full-colour 3D holographic augmented-reality displays with metasurface waveguides. *Nature* **2024**, *629*, 791−797.

(9) Draper, C. T.; Blanche, P.-A. Holographic curved waveguide combiner for HUD/AR with 1-D pupil expansion. *Opt. Express* **2022**, *30*, 2503−2516.

(10) Bang, K.; Jang, C.; Lee, B. Curved holographic optical elements and applications for curved see-through displays. *J. Inf. Disp.* **2019**, *20*, 9−23.

(11) Ni, X.; Wong, Z. J.; Mrejen, M.; Wang, Y.; Zhang, X. An ultrathin invisibility skin cloak for visible light. *Science* **2015**, *349*, 1310−1314.

(12) Khan, M. S.; Shakoor, R. A.; Fayyaz, O.; Ahmed, E. M. A focused review on techniques for achieving cloaking effects with metamaterials. *Optik* **2024**, *297*, 171575.

(13) Huang, Q.; Gan, L. T.; Fan, J. A. Conformal Volumetric Grayscale Metamaterials. *Adv. Mater.* **2023**, *35* (12), 2204688.

(14) Hecht, E. *Optics*, 5th ed.; Pearson Education, Inc.: Boston, 2017.

(15) Sasián, J. *Introduction to Lens Design*; Cambridge University Press, 2019.

(16) Yu, N.; Capasso, F. Flat optics with designer metasurfaces. *Nat. Mater.* **2014**, *13*, 139−150.

(17) Ni, X.; Kildishev, A. V.; Shalaev, V. M. Metasurface holograms for visible light. *Nat. Commun.* **2013**, *4*, 2807.

(18) Jiang, Q.; Jin, G.; Cao, L. When metasurface meets hologram: principle and advances. *Adv. Opt. Photonics* **2019**, *11*, 518−576.

(19) Balthasar Mueller, J. P.; Rubin, N. A.; Devlin, R. C.; Groever, B.; Capasso, F. Metasurface Polarization Optics: Independent Phase Control of Arbitrary Orthogonal States of Polarization. *Phys. Rev. Lett.* **2017**, *118*, 113901.
(20) Cohen, E.; Larocque, H.; Bouchard, F.; Nejadsattari, F.; Gefen, Y.; Karimi, E. Geometric phase from Aharonov−Bohm to Pancharatnam−Berry and beyond. *Nat. Rev. Phys.* **2019**, *1*, 437−449.
(21) Zhang, X.; Li, Q.; Liu, F.; Qiu, M.; Sun, S.; He, Q.; Zhou, L. Controlling angular dispersions in optical metasurfaces. *Light: Sci. Appl.* **2020**, *9* (1), 76.
(22) Wu, C.; Arju, N.; Kelp, G.; Fan, J. A.; Dominguez, J.; Gonzales, E.; Tutuc, E.; Brener, I.; Shvets, G. Spectrally selective chiral silicon metasurfaces based on infrared Fano resonances. *Nat. Commun.* **2014**, *5*, 3892.
(23) Leitis, A.; Tittl, A.; Liu, M.; Lee, B. H.; Gu, M. B.; Kivshar, Y. S.; Altug, H. Angle-multiplexed all-dielectric metasurfaces for broadband molecular fingerprint retrieval. *Sci. Adv.* **2019**, *5* (5), No. eaaw2871.
(24) Sell, D.; Yang, J.; Doshay, S.; Yang, R.; Fan, J. A. Large-Angle, Multifunctional Metagratings Based on Freeform Multimode Geometries. *Nano Lett.* **2017**, *17*, 3752−3757.
(25) Jiang, J.; Fan, J. A. Multiobjective and categorical global optimization of photonic structures based on ResNet generative neural networks. *Nanophotonics* **2021**, *10*, 361−369.
(26) Chen, M.; Jiang, J.; Fan, J. A. Design Space Reparameterization Enforces Hard Geometric Constraints in Inverse-Designed Nanophotonic Devices. *ACS Photonics* **2020**, *7*, 3141−3151.
(27) Wang, C.; Chen, N.; Heidrich, W. dO: A Differentiable Engine for Deep Lens Design of Computational Imaging Systems. *Ieee Transactions On Computational Imaging* **2022**, *8*, 905−916.
(28) Goodman, J. *Introduction to Fourier Optics*, 4t. ed.; W. H. Freeman, 2017.
(29) Taflove, A.; Hagness, S. C. *Computational Electrodynamics: the Finite-Difference Time-Domain Method*, 3rded.; Artech House, 2005.
(30) Yu, N.; Genevet, P.; Kats, M. A.; Aieta, F.; Tetienne, J.-P.; Capasso, F.; Gaburro, Z. Light Propagation with Phase Discontinuities: Generalized Laws of Reflection and Refraction. *Science* **2011**, *334*, 333−337.
(31) Cheng, H.-H.; Leportier, T.; Huynh, D.-N.; Niegemann, J.; Reid, A.; Chen, W.-H. *Ray-tracing method for large-scale metalenses in multiwavelength imaging system*; SPIE, 2025; 42−48.
(32) Zhu, Z.; Liu, Z.; Zheng, C. Metalens enhanced ray optics: an end-to-end wave-ray co-optimization framework. *Opt. Express* **2023**, *31*, 26054−26068.
(33) Zhang, Q.; Lin, P.; Yu, Z.; Zhang, C.; Liu, Y.; Wang, M.; Fan, Q.; Wang, C.; Xu, T.; Zheng, Z. Vectorial Generalized Snell's Law-Enabled Differentiable Ray Tracing for Large-Aperture Visible Achromatic Hybrid Meta-Optics. *Laser Photonics Rev.* **2025**, *19* (24), No. e00448.
(34) Shi, J.; Li, D.; Zhang, J.; Wei, X.; Nie, Y. Unified ray-wave model for end-to-end imaging in refractive−diffractive hybrid optics. *Opt. Express* **2026**, *34*, 2296−2310.
(35) Shih, K.-H.; Renshaw, C. K. Hybrid meta/refractive lens design with an inverse design using physical optics. *Appl. Opt.* **2024**, *63*, 4032−4043.
(36) Yang, X.; Souza, M.; Wang, K.; Chakravarthula, P.; Fu, Q.; Heidrich, W. *End-to-End Hybrid Refractive-Diffractive Lens Design with Differentiable Ray-Wave Model*; SIGGRAPH Asia, 2024; 1−11.
(37) Mout, M.; Flesch, A.; Wick, M.; Bociort, F.; Petschulat, J.; Urbach, P. Ray-based method for simulating cascaded diffraction in high-numerical-aperture systems. *J. Opt. Soc. Am. A* **2018**, *35*, 1356−1367.
(38) Ellepola, K. H.; Rajapaksha, T. D.; Remley, E. E.; Nguyen, M. L. P.; Macdonnell, D. G.; Leckey, J. P.; Vinh, N. Q. Monte Carlo ray-tracing simulations for diffractive optics. *Opt. Express* **2026**, *34*, 4465−4480.
(39) Li, T.-M.; Aittala, M.; Durand, F.; Lehtinen, J. Differentiable Monte Carlo ray tracing through edge sampling. *ACM Trans. Graph* **2018**, *37* (6), 1−11.
(40) Steinberg, S.; Ramamoorthi, R.; Bitterli, B.; d'Eon, E.; Yan, L.-Q.; Pharr, M. A Generalized Ray Formulation for Wave-Optics Rendering. *arXiv*, **2024**, .
(41) Tokdar, S. T.; Kass, R. E. Importance sampling: a review. *WIREs Comput. Stat.* **2010**, *2*, 54−60.
(42) Bengio, Y.; Léonard, N.; Courville, A. Estimating or Propagating Gradients Through Stochastic Neurons for Conditional Computation. *arXiv*, **2013**, .
(43) Yin, P.; Lyu, J.; Zhang, S.; Osher, S.; Qi, Y.; Xin, J. Understanding Straight-Through Estimator in Training Activation Quantized Neural Nets. *arXiv*, **2019**, .
(44) Yang, X.; Fu, Q.; Heidrich, W. Curriculum learning for ab initio deep learned refractive optics. *Nat. Commun.* **2024**, *15*, 6572.
(45) Hu, X.; Xu, W.; Fan, Q.; Yue, T.; Yan, F.; Lu, Y.; Xu, T. Metasurface-based computational imaging: a review. *Adv. Photonics* **2024**, *6*, 014002.
(46) Chen, M.; Lupoiu, R.; Mao, C.; Huang, D.-H.; Jiang, J.; Lalanne, P.; Fan, J. A. High Speed Simulation and Freeform Optimization of Nanophotonic Devices with Physics-Augmented Deep Learning. *ACS Photonics* **2022**, *9*, 3110−3123.
(47) Mao, C.; Lupoiu, R.; Dai, T.; Chen, M.; Fan, J. A., Towards General Neural Surrogate Solvers with Specialized Neural Accelerators. In *Proceedings of the 41st International Conference on Machine Learning*; PMLR, 2024.
(48) Dai, T.; Shao, Y.; Mao, C.; Wu, Y.; Azzouz, S.; Zhou, Y.; Fan, J. A. Shaping freeform nanophotonic devices with geometric neural parameterization. *Npj Comput. Mater.* **2025**, *11* (1), 259.
(49) Lupoiu, R.; Shao, Y.; Dai, T.; Mao, C.; Edée, K.; Fan, J. A. A Multi-Agentic Framework for Real-Time, Autonomous Freeform Metasurface Design. *Sci. Adv.* **2025**, *11* (44), No. eadx8006.
(50) Mao, C.; Fan, J. A. Accurate and scalable deep Maxwell solvers. *Proc. Natl. Acad. Sci. U. S. A.* **2026**, *123* (18), No. e2530330123.